\documentclass[12pt]{iopart}

\begin{document}

\title[Sum rules for the hydrogen atom]
{On certain sum rules for the hydrogen atom}

\author{Francisco M. Fern\'{a}ndez}

\address{INIFTA (UNLP, CONICET), Divisi\'on Qu\'imica
Te\'orica, Blvd. 113 S/N,  Sucursal 4, Casilla de Correo 16, 1900
La Plata, Argentina}

\ead{fernande@quimica.unlp.edu.ar}

\maketitle

\begin{abstract}
We show that some sum rules for the hydrogen atom derived recently
are incorrect because the authors did not take into account the
continuous part of the spectrum in the sum over intermediate
states.
\end{abstract}

\section{Introduction}

\label{sec:intro}

Chair and Dalabeeh\cite{CD05} and more recently Chair et
al\cite{CJSSR11} derived sum rules for the hydrogen atom by means
of the Rayleigh--Schr\"{o}dinger perturbation theory. To this end,
they compared the explicit sum over intermediate states with the
exact result provided by the method of Dalgarno and
Lewis\cite{DL55}.

The purpose of this paper is the analysis of those sume rules. In
section~\ref{sec:PT} we outline the method of Dalgarno and Lewis
and derive a general sum rule. In section~\ref{sec:sum_rule} we
discuss the validity of the sum rules derived by Chair and
Dalabeeh\cite{CD05} and Chair et al\cite {CJSSR11}. Finally, in
section~\ref{sec:conclusions} we draw conclusions.

\section{Perturbation theory}

\label{sec:PT}

Suppose that we split the hamiltonian operator $H=H_{0}+H^{\prime }$ into
the unperturbed $H_{0}$ and perturbation parts $H^{\prime }$ as is customary
in Rayleigh--Schr\"{o}dinger perturbation theory. We assume that we can
solve the eigenvalue equation for $H_{0}$
\begin{equation}
H_{0}\left| n\right\rangle =E_{n}^{(0)}\left| n\right\rangle ,\;n=0,1,\ldots
\label{eq:Schro_H0}
\end{equation}
where $E_0 < E_1 \leq E_2 \leq \ldots$ The first and second order
perturbation corrections to the ground--state energy (assumed to
be nondegenerate) are
\begin{eqnarray}
E_{0}^{(1)} &=&\left\langle 0\right| H^{\prime }\left| 0\right\rangle
\nonumber \\
E_{0}^{(2)} &=&\sum_{n>0}\frac{\left| \left\langle n\right| H^{\prime
}\left| 0\right\rangle \right| ^{2}}{E_{0}^{(0)}-E_{n}^{(0)}}
\label{eq:pert_corr}
\end{eqnarray}
where
\begin{equation}
\sum_{n=0}\left| n\right\rangle \left\langle n\right| =\hat{1}
\label{eq:complete}
\end{equation}
is the identity operator. If $H_{0}$ exhibits continuous spectrum
then the sums above should include the corresponding integrals
over such states\cite {DL55}.

The method of Dalgarno and Lewis\cite{DL55} enables one to obtain the sum in
equation (\ref{eq:pert_corr}) in closed form. If we can find an operator $F$
such that
\begin{equation}
H^{\prime }\left| 0\right\rangle =[H_{0},F]\left| 0\right\rangle
\end{equation}
then we obtain the sum rule
\begin{equation}
\sum_{n>0}\frac{\left| \left\langle n\right| H^{\prime }\left|
0\right\rangle \right| ^{2}}{E_{0}^{(0)}-E_{n}^{(0)}}=\left\langle 0\right|
H^{\prime }F\left| 0\right\rangle -\left\langle 0\right| H^{\prime }\left|
0\right\rangle \left\langle 0\right| F\left| 0\right\rangle
\label{eq:gen_sum_rule}
\end{equation}
provided that the set of eigenfunctions satisfies
(\ref{eq:complete}).

\section{The sum rules for hydrogen}

\label{sec:sum_rule}

In order to obtain their sum rules Chair and Dalabeeh\cite{CD05}
and Chair et al\cite{CJSSR11} proved that $\left\langle 0\right|
H^{\prime }\left| 0\right\rangle =0$ and managed to calculate the
matrix elements $ \left\langle n\right| H^{\prime }\left|
0\right\rangle $ and $\left\langle 0\right| H^{\prime }F\left|
0\right\rangle $ that appear in equation (\ref {eq:gen_sum_rule}).
However, since they omitted the continuous spectrum of hydrogen
the bound states in their sums do not span the whole state space
(that is to say: do not satisfy equation (\ref {eq:complete}))
and, consequently, their sum rules cannot be exact.

We first analyse the sum rule derived by Chair et al\cite{CJSSR11}
\begin{equation}
S=2^{10}\sum_{n=3}^{\infty
}n^{7}(n^{2}-4)\frac{(n-1)^{2n-6}}{(n+1)^{2n+6}}= \frac{15}{2}
\label{eq:sum_rule}
\end{equation}
If we take into account that
\begin{equation}
\frac{(n-1)^{2n-6}}{(n+1)^{2n+6}}<\frac{1}{(n+1)^{12}}
\end{equation}
then we derive the inequality

\begin{eqnarray}
S<2^{10}\sum_{n=3}^{\infty }\frac{n^{7}(n^{2}-4)}{(n+1)^{12}}
&=&-19456\zeta {\left( 11\right) }-57344\zeta {\left( 9\right)
}+43008\zeta {\left( 7\right) } \nonumber\\
&&+32768\zeta {\left(5\right) }+1024\zeta {\left( 3\right)
}+\frac{707584\pi ^{12}}{212837625}\nonumber \\
&&+\frac{16384\pi ^{10}}{31185}+\frac{1024\pi
^{8}}{675}-\frac{8192\pi ^{6}}{ 135}-\frac{512\pi
^{4}}{5}+\frac{3}{4}\nonumber \\
&\approx &6.889304238
\end{eqnarray}
where $\zeta (s)=\zeta (s,1)$ and
\begin{equation}
\zeta (s,q)=\sum_{j=0}^{\infty }\frac{1}{(j+q)^{s}}
\end{equation}
is the zeta function. We appreciate that $S<6.889304238$ is
considerably smaller than the result (\ref{eq:sum_rule}) given by
Chair et al \cite {CJSSR11}. These authors compared their sum rule
with one of those derived earlier by Bell\cite{B67}. However, Bell
obtained the results of the sums and did not calculate the terms
in the series explicitly. Chair et al\cite{CJSSR11} did it but
only for the discrete states; for this reason their explicit sum
rule is incorrect.

Chair and Dalabeeh\cite{CD05} derived the following sum rule
\begin{equation}
S=\frac{2^{8}}{3}\sum_{n=2}^{\infty }n^{5}\frac{(n-1)^{2n-4}}{(n+1)^{2n+4}}=1
\label{eq:sum_rule_2}
\end{equation}
In this case we slightly modify the strategy applied to the
previous example. Note that
\begin{eqnarray}
\frac{2^{8}}{3}\sum_{n=m+1}^{\infty }n^{5}\frac{(n-1)^{2n-4}}{(n+1)^{2n+4}}
&<&\frac{2^{8}}{3}\sum_{n=m+1}^{\infty }\frac{n^{5}}{(n+1)^{8}}=  \nonumber
\\
&=&-\frac{256\zeta {(8,}m{+2)}}{3}+\frac{1280\zeta
{(7,}m{+2)}}{3}-\frac{
2560\zeta {(6,}m{+2)}}{3}  \nonumber \\
&&+\frac{2560\zeta {(5,}m{+2)}}{3}-\frac{1280\zeta
{(4,}m{+2)}}{3}\nonumber \\
&&+\frac{ 256\zeta {(3,}m{+2)}}{3}
\end{eqnarray}
and that
\begin{equation}
S<S_{m}=\frac{2^{8}}{3}\sum_{n=2}^{m}n^{5}\frac{(n-1)^{2n-4}}{(n+1)^{2n+4}}+
\frac{2^{8}}{3}\sum_{n=m+1}^{\infty }\frac{n^{5}}{(n+1)^{8}}
\end{equation}
where, obviously, $S_{m+1}<S_{m}$. Since $S_{7}=0.9461580468$
prove that $S$ is smaller than unity.

\section{Conclusions}

\label{sec:conclusions}

We have proved that the sum rules derived by Chair and
Dalabeeh\cite{CD05} and Chair et al\cite{CJSSR11} are incorrect
because the authors did not take into account the continuous
spectrum of the hydrogen--atom Hamiltonian $H_{0}$. Although those
authors may have calculated the result of the sum exactly by means
of the method of Dalgarno and Lewis\cite{DL55} they omitted the
contribution of the continuous spectrum in the sum over
intermediate states (\ref{eq:gen_sum_rule}). Therefore, their sums
are always smaller than the exact results because
\[
\sum_{n>0}^{discrete}\frac{\left| \left\langle n\right| H^{\prime
}\left| 0\right\rangle \right|
^{2}}{E_{n}^{(0)}-E_{0}^{(0)}}<\left\langle 0\right| H^{\prime
}\left| 0\right\rangle \left\langle 0\right| F\left|
0\right\rangle -\left\langle 0\right| H^{\prime }F\left|
0\right\rangle
\]

\end{document}